\documentclass[twocolumn]{Trivedi_Papers}
\usepackage{graphicx}
\usepackage[]{cite}
\usepackage{floatrow}
\usepackage{sidecap}

\usepackage{notes2bib}

\usepackage[switch]{lineno}

\usepackage{zref-xr}
\zxrsetup{%
  tozreflabel=false, 
  toltxlabel=true, 
}

\usepackage{color}

\newcommand{\green}[1]{\textcolor{cyan}{#1}}

\makeatletter
\def\blfootnote{\gdef\@thefnmark{}\@footnotetext}
\makeatother

\newcommand{\blist}{
 \begin{list}{$\bullet$}
  { \setlength{\itemsep}{2pt}
     \setlength{\parsep}{0pt}
     \setlength{\topsep}{6pt}
     } }
\newcommand{\elist}{
  \end{list}  }

\begin{document}

\title{Imaging the Beating Heart with Macroscopic Phase Stamping}
\vspace{-8pt}
\author{{\authorfont\authorcolor
Vikas Trivedi$^{1,2,3}$, Sara Madaan$^{2}$, Daniel B. Holland$^{2}$, Le A. Trinh$^{2}$, Scott E. Fraser$^{2,*}$, Thai V. Truong$^{2,*}$}\\[6pt]
\medskip
\begin{center}
    \begin{minipage}{0.8\textwidth}
\bfseries\sffamily{We present a novel approach for imaging the beating embryonic heart, based on combining two independent imaging channels to capture the full spatio-temporal information of the moving 3D structure. High-resolution, optically-sectioned image recording is accompanied by simultaneous acquisition of low-resolution, whole-heart recording, allowing the latter to be used in post-acquisition processing to determine the macroscopic spatio-temporal phase of the heart beating cycle. Once determined, or “stamped”, the phase information common to both imaging channels is used to reconstruct the 3D beating heart. We demonstrated our approach in imaging the beating heart of the zebrafish embryo, capturing the entire heart over its full beating cycle, and characterizing cellular dynamic behavior with sub-cellular resolution.}\\
           \end{minipage}
\end{center}
\vspace{-8pt}
}

\maketitle

\blfootnote{\affilfont$^\textsf{1}$Division of Biology and Biological Engineering, California Institute of Technology, Pasadena, California 91125, United States.
$^\textsf{2}$Translational Imaging Center, Department of Molecular and Computational Biology, University of Southern California, Los Angeles CA 90089, United States.
$^\textsf{3}$Current affiliations: EMBL Barcelona, C/ Dr. Aiguader 88, 08003 Barcelona, Spain \textit{and} EMBL Heidelberg, Developmental Biology Unit, 69117 Heidelberg, Germany.
{\bfseries\sffamily *Authors for correspondence: tvtruong@usc.edu or sfraser@provost.usc.edu }
}

\runningpagewiselinenumbers

\begin{figure*}[ht]
\includegraphics[width=7.15in]{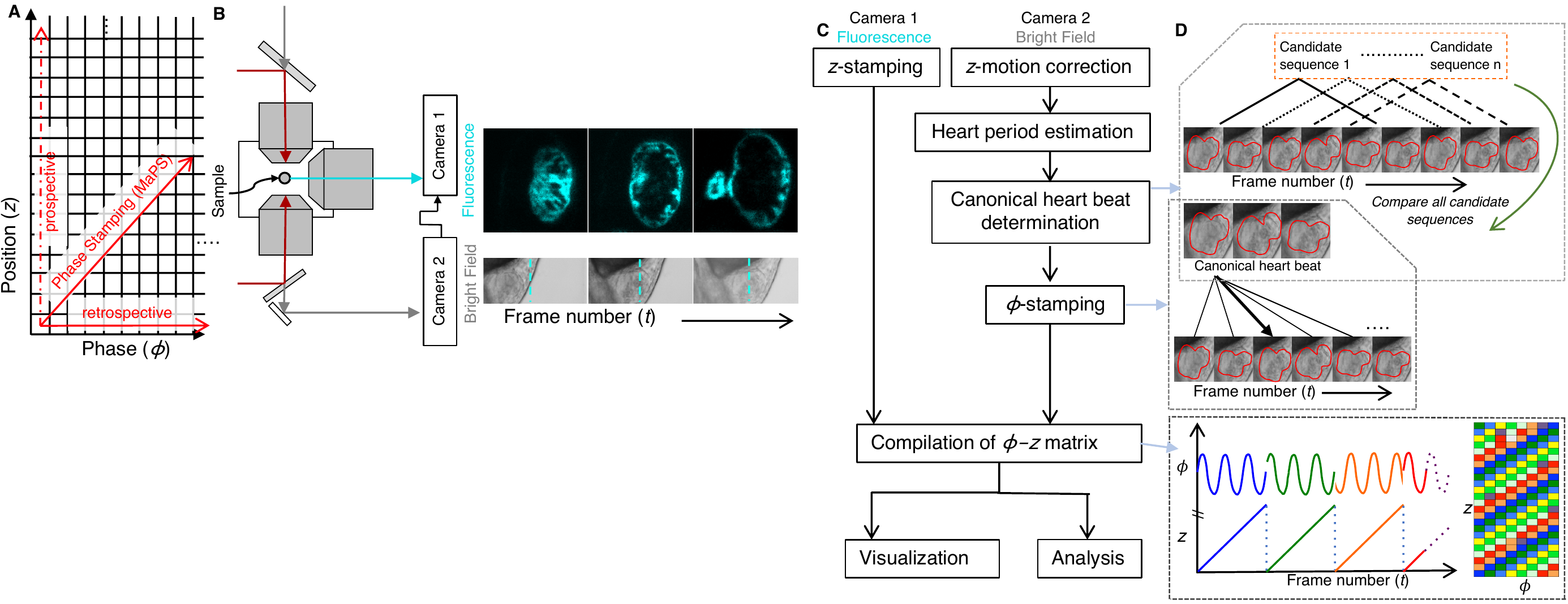}
 \caption{{\fignumfont Macroscopic Phase Stamping (MaPS) for 4D imaging of beating heart}  \textbf{(A)} 4D data acquisition of the beating heart is a systematic acquisition of 2D images at all positions ($z$) within the heart and at all spatio-temporal configuration (phase, $\phi$) of the periodically beating heart, to fill the phase-position ($\phi-z$) matrix. The desired levels of phase and spatial depth resolution will determine the size of the $\phi-z$ matrix, with upper bounds dictated by instrumentation limitations on imaging speed and depth resolution. \textit{Retrospective approaches} image all phases at one $z$-position and repeat this for the next $z$-position. \textit{Prospective approaches} image all positions one at a time at a given phase and repeat this for the next phase. Macroscopic Phase Stamping (MaPS) images the sample through continuous movement in $z$ while the heart is beating i.e. also changing its phase. 
\textbf{(B)} Schematic of the light sheet microscopy set up used for MaPS. For the fluorescent channel (\textit{cyan}), scanned light sheet (\textit{deep red}) is used for bi-directional illumination and the signal from single optical plane is detected orthogonally. For the bright-field channel (\textit{grey}), far-red LED light (\textit{grey}) illuminates the sample through one of the illumination objectives and the transmitted light is collected through the other one onto a camera synchronized to the fluorescent camera, thereby providing a macroscopic image of the heart. As the stage translates, high-resolution images over $z$ are acquired simultaneously with the bright field images. 
\textbf{(C,D)} Overview of the MaPS image reconstruction pipeline. Determination of $z$ position ($z$-stamping) is achieved by counting the frame number of fluorescent images programmed to be acquired at 1 fluorescent image per $\mu m$ movement of the stage. In the acquired bright-field image sequences, the motion observed due to $z$-scanning of the heart is subtracted to obtain a stationary region of interest, comprising of the beating heart. These images are used to estimate heart period and the number of distinct phases ($\phi$). The \textit{canonical heartbeat} is determined by comparing all possible candidate image sequences (comprising of $\phi$ frames) to obtain the image sequence most similar to all other sequences in the tested data. The images within this  \textit{canonical heartbeat} then define the spatio-temporal phases of the heartbeat. Comparing every frame in the bright field image to this \textit{canonical heartbeat} provides the phase of each frame ($\phi$-stamping). The $z-$ and $\phi$-stamps are then relayed to the corresponding fluorescent images and compiled into the $\phi-z$ matrix. The assembled 4D data is ready for visualization and analysis. 
\label{Figure1-MaPS}}
\end{figure*}

The embryonic heart undergoes critical three-dimensional spatial patterning while it is beating and pumping blood, thus valuable insights to the mechanisms of cardiac development could be gained through imaging tools that can capture both the structure and motion of the 4D (3 spatial dimensions plus time) organ. Considerable research effort has been directed towards the goal of understanding heart development, focusing on the zebrafish embryonic heart due to its small size, optical and genetic accessibility, and its  developmental similarities to other vertebrates\green{\cite{stainier2001zebrafish, harvey2002patterning, hove2003intracardiac, bartman2004early, forouhar2006embryonic, santhanakrishnan2011fluid}}. Even with the optically accessible zebrafish model system, 4D imaging of the heart faces a fundamental challenge: most optical imaging modalities are intrinsically 2D in nature, i.e. the detector captures a single 2D image slice at a time, thus it is impossible to directly acquire the 4D data-cube of the beating heart. Conceptually, illustrated in Fig.\ref{Figure1-MaPS}A, the 4D data acquisition can be represented as the systematic collection of 2D images at all depth positions ($z$) that encompass the organ, and at all spatio-temporal configuration (phase, $\phi$) of the periodic beating motion of the heart. Data acquisition into the $\phi-z$ matrix can only be captured one entry at a time, thus presenting an intrinsic limitation to direct 4D data capture.

Two general strategies have been devised to address the challenge of optical 4D imaging of the beating heart (Fig.\ref{Figure1-MaPS}A), both with unique advantages and limitations.  The \textit{prospective} approach uses an independent information channel to monitor the heart in real time, and triggers 2D image acquisition at a specific phase within the cardiac cycle (e.g. at peak ventricular expansion), capturing one image for each beating cycle\green{\cite{taylor2011real, lee2012real, taylor2019adaptive}}. This approach excels at capturing high-resolution images of the heart at a particular phase in its beating cycle, but does not provide information on the full 4D beating motion of the heart throughout the cardiac cycle, thus precluding its utility in studies where the beating motion itself is important. The \textit{retrospective} approach for imaging the heart acquires continuous 2D movies of the beating organ at each $z$-position sequentially, then in post-processing uses the assumptions of cardiac periodicity and sample spatial continuity across the $z$-axis to synchronize the phase-sequence of the movies and populate the $\phi-z$ matrix \green{\cite{forouhar2006embryonic, liebling2005four, liebling2006rapid, liebling2006nonuniform, mahou2014multicolor, mickoleit2014high, staudt2014high, trivedi2015dynamic}}, to yield the full 4D beating motion of the heart throughout the cardiac cycle. The imposition of sample spatial continuity across the $z$-axis potentially can lead to artefacts when this assumption is not valid \green{\cite{mickoleit2014high}}, such as when there is significant signal variation along the $z$-axis, e.g. stemming from contraction or calcium waves that propagate through the heart as part of the mechanisms that govern the beating.

Here we present a novel approach for 4D imaging of the beating heart that combines elements from both the prospective and retrospective methods to overcome their respective limitations. In our approach, we employ an additional imaging path to obtain a low-resolution bright-field image of the beating heart, captured synchronously with the optically-sectioned fluorescent images (Fig.\ref{Figure1-MaPS}B), so that for every high-resolution fluorescent image we have a bright-field image that contains the macroscopic spatio-temporal information about phase ($\phi$) of the beating heart. Data acquisition consists of continuously scanning the sample in $z$, and recording the continuous image sequences from the synchronously-triggered fluorescent and bright-field cameras, effectively filling in the $\phi-z$ matrix diagonally (Fig.\ref{Figure1-MaPS}A). Post-acquisition, we use the bright-field images to quantitatively determine the instantaneous phase of heart for each acquired fluorescent image, a process that we term macroscopic phase-stamping (MaPS). Coupled with the known $z$-position of every fluorescent image (i.e. $z$-stamping), we then have both the $\phi$ and $z$ values to assemble the optically sectioned fluorescent images into the $\phi-z$ matrix (Fig.\ref{Figure1-MaPS}A), achieving the full 4D description of the beating heart.

\begin{figure*}[ht]
\centering
\includegraphics[width=7.15in]{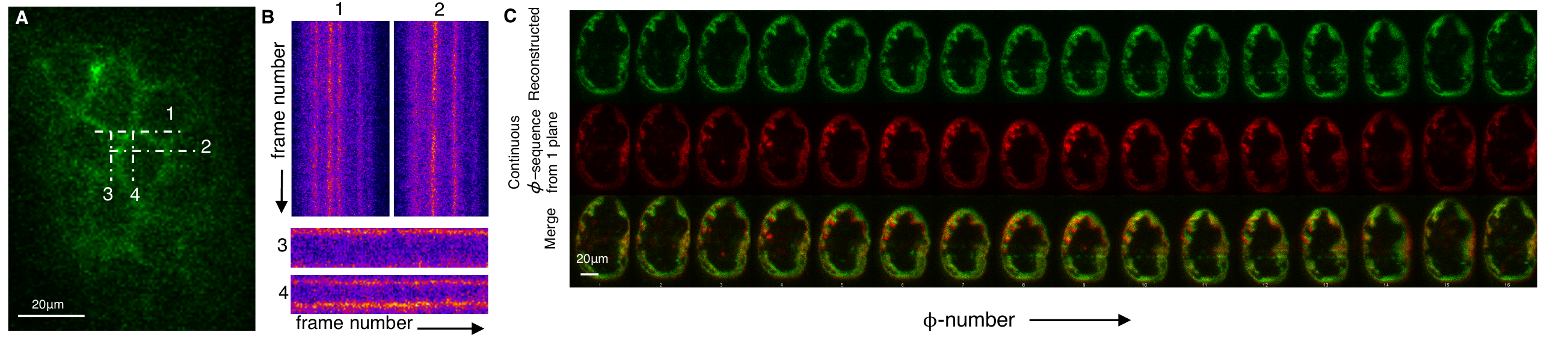}
 \caption{{\fignumfont Accuracy of $z-$ and phase-stamping} \textbf{(A)}  Single z-section showing the cells of the static pericardial wall, reconstructed through MaPS. Zebrafish sample is the transgenic line  $Gt(ctnna-citrine)^{ct3a/ct3a}$, where the fluorescence label highlights the cellular boundaries. The numbered lines denote the section used for obtaining the kymographs across all phases in (B).
\textbf{(B)} Kymographs showing the signal along the numbered lines shown in (A), over all the phases in the 4D reconstruction. As the pericardial wall is a static structure which remains stationary during the heartbeat (i.e. all phases are equivalent), kymographs across the pericardial cellular boundaries are expected to be continuous lines only in the case of accurate $z$-stamping, as were observed here. 
\textbf{(C)} Comparison of sequence of images (across phases from a single $z$-section) between the image series reconstructed with MaPS (\textit{green}) and image series directly acquired by parking the light sheet at the same position without any stage motion (\textit{red}). Sample is the transgenic line $Gt(tpm4-citrine)^{ct31a/+}$, where the fluorescence label highlights the cytoplasm of myocardial cells. The high degree of spatio-temporal overlap between the two image sequences indicate that accurate $z-$ and $\phi-$stamping are obtained with MaPS.
\label{Figure2-Accuracy}}
\end{figure*}

The advantage of our MaPS approach comes from the independent and simultaneous acquisition of both high-resolution, local information (the optically-sectioned fluorescent images) and low-resolution, global information (the bright field images) of the heart. In reconstructing the 4D beating of the heart, we invoke the putative periodicity of the heart motion, but we have the independent information to quantitatively gauge the validity of this assumption. We achieve the combined benefits of both the prospective and retrospective approaches, providing the full 4D capture of the heart beating motion, without needing to assume sample continuity along the $z$-axis. Our MaPS 4D imaging approach relies on information at both local and global scales to provide an accurate statistical-ensemble description of the 4D beating motion of the heart.

We implemented the MaPS 4D imaging approach with a custom-built light sheet microscope, capturing the 4D beating of the embryonic zebrafish heart at subcellular resolution, at up to 5 days post fertilization (dpf). Light sheet microscopy\green{\cite{huisken2004optical}} also known as Selective Plane Illumination Microscopy (SPIM), illuminates the sample one optical section at a time from the side, providing enhanced contrast while reducing photo-induced sample damage, and thus is well suited for imaging the live beating embryonic heart\green{\cite{taylor2011real, lee2012real, mahou2014multicolor, mickoleit2014high, trivedi2015dynamic}}. Our SPIM setup\green{\cite{trivedi2015dynamic, truong2011deep}}  employed scanned light sheets to provide the optically-sectioned subcellular-resolution fluorescence imaging of the beating heart (Fig.\ref{Figure1-MaPS}B, Methods, Supplementary Figs. 1-3). Two-photon light sheet excitation was used to improve depth penetration and contrast, especially for zerbrafish at later developmental stages\green{\cite{trivedi2015dynamic}}, but our method should work as well with one-photon excitation. We added another imaging path, orthogonal  to the fluorescence path, to provide a low-magnification, bright-field imaging channel, allowing us to capture the macroscopic spatio-temporal phases of the beating heart for the phase-stamping (Fig.\ref{Figure1-MaPS}B). Data collection was carried out with custom control software where both the fluorescence and bright-field imaging channel were acquired synchronously, as the sample is repeatedly scanned across the illuminating light sheets, collecting information into the $\phi-z$ matrix along the diagonals as depicted schematically in Fig.\ref{Figure1-MaPS}A (Methods, Supplementary Figs. 2-3).

The overall workflow of the image reconstruction pipeline is depicted in Fig.\ref{Figure1-MaPS}C, and described in details in Supplementary Methods (Supplementary Fig. 3). Briefly, it starts with the determination of $z$ position ($z$-stamping) of each pair of the acquired fluorescent and bright-field images, by matching the frame number to the hardware signal that was used to drive the $z$-stage. The $z$-stamps are then used to subtract out the sample motion due to the $z$-scanning in the bright-field movie sequence, yielding a stationary region of interest of the beating heart. The motion-corrected movie is then used to determine the average heart period and the number of distinct phases ($\phi$). Next, using image correlation we compare heartbeats to each other to obtain the \textit{canonical heartbeat}, defined as the image sequence most similar to all other sequences (Fig. 1C, D, Methods). As the heart in general exhibits natural variability\green{\cite{moorman2011cardiovascular, li2014robust}} in its beating period and motion, the canonical heartbeat represents the most suitable standard to which we can compare all other heartbeats. The sequence of images within the canonical heartbeat then defines the spatio-temporal phases of the beat. Comparing every frame in the bright field image to this canonical heartbeat we determine the phase of each frame ($\phi$-stamping) (Fig. 1C, D). The phase-stamps of the bright-field images are then linked to the corresponding fluorescent images, each of which now has both its $z$-stamp and $\phi$-stamp. We then compile the images into the $\phi-z$ matrix, and the assembled 4D data is ready for visualization and analysis. With the raw 2D imaging speed rate at 80 frames/s, the MaPS protocol could identify around 16-20 distinct phases, depending upon the developmental stage of the sample. With the heartbeat frequency of around 2.5-3 Hz, the number of identified phases translated to a volumetric rate for MaPS reconstructions of around 50 volumes/s. To assess the accuracy and robustness of MaPS, we used the stationary epicardial wall to confirm the $z$-stamping, and we compared the reconstructed phases with single-plane movies to confirm the $\phi$-stamping (Fig.\ref{Figure2-Accuracy}).

\begin{figure}[ht]
\centering
\includegraphics[width=3.5in]{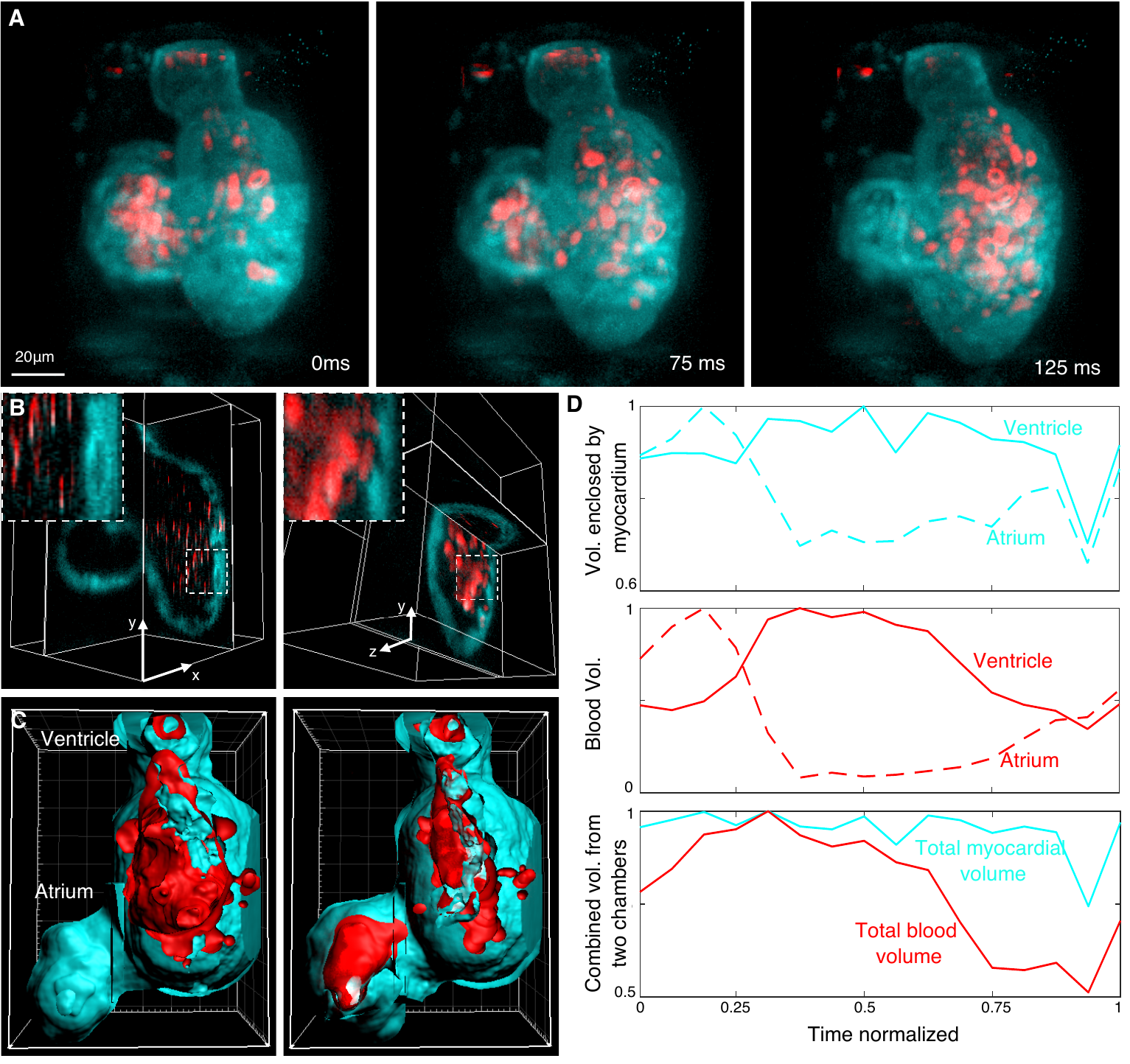}
 \caption{{\fignumfont Visualization of myocardial volume changes and blood flow in the zebrafish beating heart captured with MaPS} \textbf{(A-B)} Three time points showing 3D-rendered views in the 4D reconstructed image series of the beating heart of the double transgenic line $Gt(tpm4-citrine)^{ct31a/+};Tg(gata5:dsRed)$ at 4.5 days post fertilization (dpf). Cytoplasmic Citrine (\textit{cyan}) and dsRed (\textit{red}) show myocardial cells and red blood cells, respectively. 
\textbf{(C)} Digitally cropped iso-surface renderings of myocardium and the volume of the blood showing their relative volumes at 2 different phases during the beating.
\textbf{(D)} Normalized volume changes over the heart beating cycle, shown for the blood volume (\textit{red}) and the myocardial space (\textit{cyan}) for the two chambers, atrium and ventricle, both separately and together.
  \label{Figure3-Flow}}
\end{figure}

We applied MaPS to capture the dynamic motion of the myocardium, at subcellular resolution, throughout its beating cycle, along with the blood flow in 4.5-5.5 dpf zebrafish embryos (Fig.\ref{Figure3-Flow}A-D, Fig. S4, Movies 1-3). The entire 3D structure of the two cardiac chambers, the atrium and ventricle, was captured, with the moving myocardial walls and cardiac lumen clearly delineated at all phases. From the reconstructed 4D images, we quantified the volume of the two chambers during the cardiac cycle (Fig.\ref{Figure3-Flow}E-F), finding an out-of-phase temporal relationship between the ventricular and atrial volumes, as expected for the sequential blood flow from the atrium to the ventricle.

\begin{figure}[ht]
\includegraphics[width=3.5in]{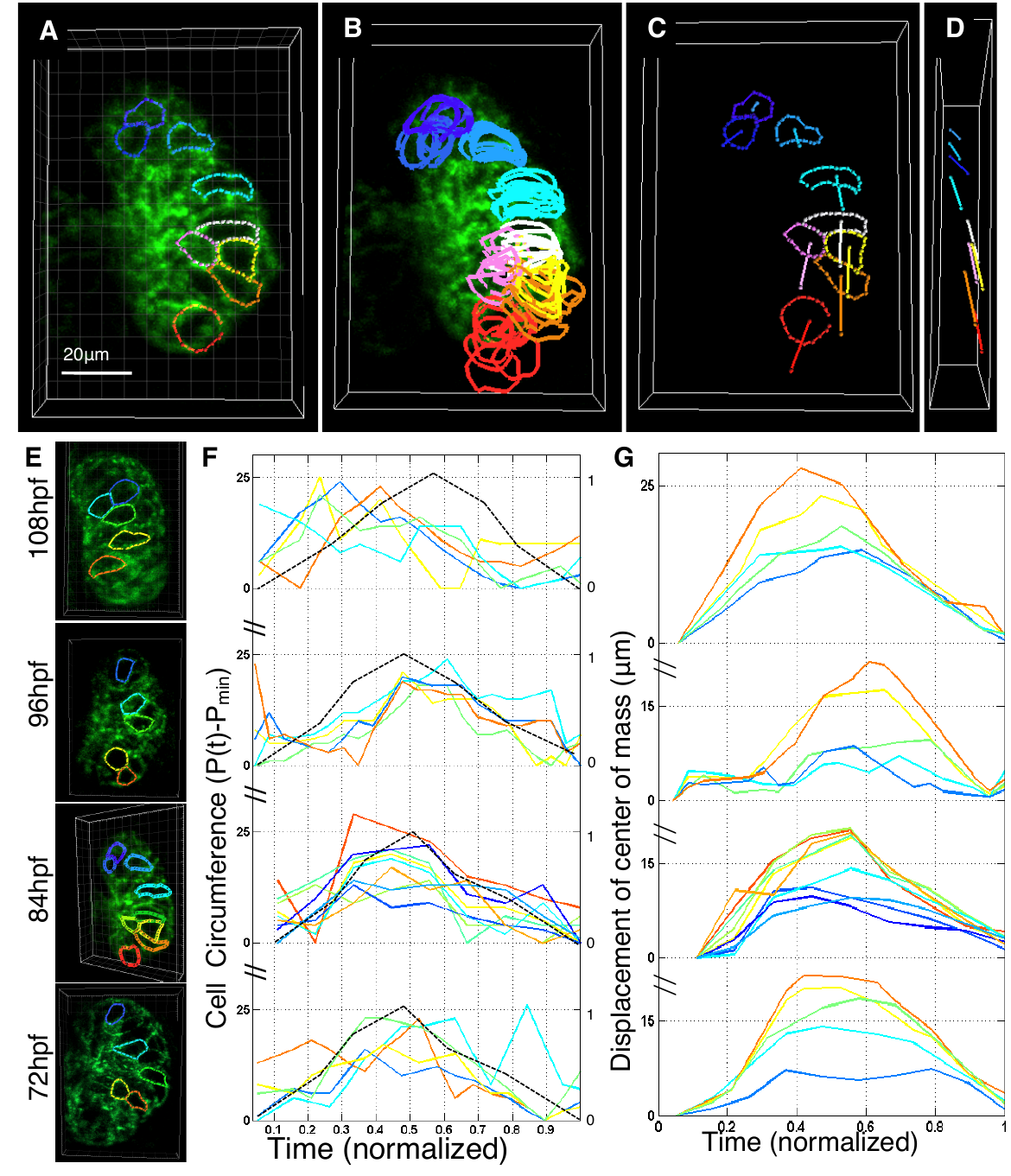}
 \caption{{\fignumfont Changes in cellular size over beating cycles in cardiomyocytes} \textbf{(A)} 3D-rendered view of a single time point in the 4D reconstructed image series of the beating ventricle at 84 hpf in the transgenic line $Gt(ctnna-citrine)^{ct3a/ct3a}$ where the fluorescence label highlights the boundaries of all cells in the developing heart. Segmentation was manually performed in Imaris by first marking distinct points (spots in the movie) along the cardiomyocytes’ boundary and then connecting them. Sufficient number of points was chosen to ensure smoothly varying representation of cell boundaries (9 representative cells shown here) at every time point. The outflow track of the heart is at the top of the image.
\textbf{(B)} Segmented cell boundaries at different phases in the heart beat shown for the same 9 representative cells in the ventricle.
\textbf{(C,D)} Segmented cell boundaries at a single phase shown for 9 representative cells along with their total displacement during the heartbeat (straight lines) from ventral (C) and lateral (D) views.
\textbf{(E)} Segmented cell boundaries at a single phase shown for representative cells in the ventricle at 4 different developmental stages starting 72 hpf.
\textbf{(F)} Changes in the cell circumference over the temporally-normalized heartbeat, shown for the representative cells (marked in E) at 4 different developmental stages. The colored lines denote changes in cellular circumference and black dotted-lines denote the overall change in the volume of the ventricle at corresponding stages.
\textbf{(G)} Displacement of the centers of mass over the temporally-normalized heartbeat, shown for the representative cells (marked in E) at 4 different developmental stages. 
\label{Figure4-Shape}}
\end{figure}

Using MaPS we also measured multiple cardiomyocytes’ behavior as they undergo contraction and expansion in $Gt(ctnna-citrine)^{ct3a/ct3a}$ zebrafish heart, where the endogenous expression of the fusion construct alpha catenin-Citrine highlights the boundaries of all cells in the developing heart (Fig.\ref{Figure4-Shape}). The high spatio-temporal resolution allowed us to follow the cyclic cellular motion of multiple representative cardiomyocytes in the ventricle over time (Fig.\ref{Figure4-Shape}A, Movies 4-7). We tracked these cells over the beating cycle to determine their displacement in 3D within the intact tissue (Fig.\ref{Figure4-Shape}B-D). Further, we segmented and quantified the nature of contraction of individual cardiomyocytes, spread throughout the ventricular tissue, over different developmental stages (Fig.\ref{Figure4-Shape}E-G). Our analysis shows that the overall contraction of the ventricle, as assessed by change in volume over time (Fig.\ref{Figure4-Shape}F, black dotted-curves), undergoes a  gradual increase in volume followed by contraction, as expected; whereas, individual cardiomyocytes (Fig.\ref{Figure4-Shape}F, colored curves) display a range of heterogeneous behavior with varying amount of time lags between each other. The centers of mass of the cardiomyocytes move through the beating cycle with strong correlation to anatomy, such that the cells near to the outflow tract, anchored to the pericardial cavity, displace less as compared to the cells farther away (Fig.\ref{Figure4-Shape}G).

In summary, we have presented MaPS as a novel approach to capture the 3D motion of the beating heart. By utilizing the low-resolution, global-information bright-field imaging channel to phase-stamp the high-resolution, local-information fluorescence channel, we can acquire a statistically accurate 4D representation of the periodic beating motion of the heart. The accuracy of MaPS is expected to be better with higher image acquisition rate. With our current implementation of MaPS, we expect to be subjected to quantization errors in $\phi-$ and $z-$position of $\pm 1$ phase points and $\pm 1 \mu m$ respectively. Strategies to achieve super-resolution in the sampling rate from extended temporal acquisition window\green{\cite{bertero2003super}} could be pursued in future work to achieve higher accuracy in the phase stamping process. In our implementation of MaPS, we have used fluorescent SPIM and bright-field imaging as the high-resolution and the phase channel, respectively, since they are well suited for the zebrafish embryonic heart model system. However, the generalized principle of MaPS, where 4D imaging is achieved by combining two independent acquisition channels that together span the entire relevant spatio-temporal range, could be implemented with a variety of imaging, and even non-imaging, data acquisition modalities, such as optical coherence tomography, ultrasound imaging, eletrocardiography, and magnetic resonance imaging\green{\cite{markl2003time, jerecic2004ecg, bartling2010gating, jenkins12vivo, yagel20073d, hung2007lang}} depending on the sample system under study. We envision the generalized strategy of MaPS of integrating local and global information will find widespread use in biomedical imaging.

\bibliographystyle{science}
\sffamily
\begin{footnotesize}
\bibliography{Ref_MaPS}
\end{footnotesize}

\begin{acknow}
We thank Leigh Ann Fletcher and Lorna McFarlane for fish care. 
 \end{acknow}
 


\begin{funding}
This work was funded by the Rosen Center for Bioengineering (Caltech), a Center of Excellence in Genomic Science grant (NIH/NHGRI P50HR004071), a FaceBase grant (NIH Grant U01DE020063) and the Translational Imaging Center at USC. 
 \end{funding}


\end{document}